\documentclass[12pt]{article}
\usepackage{epsfig,cite}
\usepackage {amsmath}
\usepackage{amssymb}
\include{epsf}
\newcount\timecount
\newcount\hours \newcount\minutes  \newcount\temp \newcount\pmhours
\hours = \time
\divide\hours by 60
\temp = \hours
\multiply\temp by 60
\minutes = \time
\advance\minutes by -\temp
\def\hour{\the\hours}
\def\minute{\ifnum\minutes<10 0\the\minutes
            \else\the\minutes\fi}
\def\clock{
\ifnum\hours=0 12:\minute\ AM
\else\ifnum\hours<12 \hour:\minute\ AM
      \else\ifnum\hours=12 12:\minute\ PM
            \else\ifnum\hours>12
                 \pmhours=\hours
                 \advance\pmhours by -12
                 \the\pmhours:\minute\ PM
                 \fi
            \fi
      \fi
\fi
}

\def\monthname{\relax\ifcase\month 0/\or January\or February\or
   March\or April\or May\or June\or July\or August\or September\or
   October\or November\or December\else\number\month/\fi}

\def\bold#1{\setbox0=\hbox{$#1$}%
     \kern-.025em\copy0\kern-\wd0
     \kern.05em\copy0\kern-\wd0
     \kern-.025em\raise.0433em\box0 }

\def\beq{\begin{equation}}
\def\eeq{\end{equation}}

\def\ga{\mathrel{\raise.3ex\hbox{$>$\kern-.75em\lower1ex\hbox{$\sim$}}}}
\def\la{\mathrel{\raise.3ex\hbox{$<$\kern-.75em\lower1ex\hbox{$\sim$}}}}
\def\gev{{\rm \, Ge\kern-0.125em V}}
\def\tev{{\rm \, Te\kern-0.125em V}}
\def\gyr{{\rm \, G\kern-0.125em yr}}

\def\gappeq{\mathrel{\rlap {\raise.5ex\hbox{$>$}}
{\lower.5ex\hbox{$\sim$}}}}
\def\lappeq{\mathrel{\rlap{\raise.5ex\hbox{$<$}}
{\lower.5ex\hbox{$\sim$}}}}
\def\Toprel#1\over#2{\mathrel{\mathop{#2}\limits^{#1}}}

\def\m12{m_{1\!/2}}
\def\bea{\begin{eqnarray}}
\def\eea{\end{eqnarray}}

\begin{document}
\begin{titlepage}
\pagestyle{empty}
\baselineskip=21pt
\rightline{\tt hep-ph/0607002}
\rightline{CERN-PH-TH/2006-110}
\rightline{UMN--TH--2509/06}
\rightline{FTPI--MINN--06/23}
\vskip 0.2in
\begin{center}
{\large{\bf What if supersymmetry breaking appears below the GUT scale?}}
\end{center}
\begin{center}
\vskip 0.2in
{\bf John~Ellis}$^1$, {\bf Keith~A.~Olive}$^{2}$ and
{\bf Pearl Sandick}$^{2}$
\vskip 0.1in

{\it
$^1${TH Division, CERN, Geneva, Switzerland}\\
$^2${William I. Fine Theoretical Physics Institute, \\
University of Minnesota, Minneapolis, MN 55455, USA}\\
}

\vskip 0.2in
{\bf Abstract}
\end{center}
\baselineskip=18pt \noindent
%%%%%%%%%%%%%%%%%%%%%%%%%%%%%%%%%%%%%%%%%%%%%%%%%%%%%%%%%%%%%%%%%%%%%

We consider the possibility that the soft supersymmetry-breaking parameters 
$m_{1/2}$ and $m_0$ of the MSSM
are universal at some scale $M_{in}$ below the supersymmetric grand unification
scale $M_{GUT}$, as might occur in scenarios where either the primordial 
supersymmetry-breaking mechanism or its communication to the observable sector
involve a dynamical scale below $M_{GUT}$. We analyze the $(m_{1/2}, m_0)$
planes of such sub-GUT CMSSM models, noting the dependences of phenomenological,
experimental and cosmological constraints on $M_{in}$. In particular, we find that the
coannihilation, focus-point and rapid-annihilation funnel regions of the GUT-scale
CMSSM approach and merge when $M_{in} \sim 10^{12}$~GeV. We discuss
sparticle spectra and the possible sensitivity of LHC measurements to the value
of $M_{in}$.

%%%%%%%%%%%%%%%%%%%%%%%%%%%%%%%%%%%%%%%%%%%%%%%%%%%%%%%%%%%%%%%%%%%%%
\vfill
\leftline{CERN-PH-TH/2006-110}
\leftline{June 2006}
\end{titlepage}
\baselineskip=18pt
%%%%%%%%%%%%%%%%%%%%%%%%%%%%%%%%%%%%%%%%%%%%%%%%%%%%%%%%%%%%%%%%%%%%%

\section{Introduction}

The primary phenomenological reason for expecting supersymmetry to appear
at the TeV scale is to ensure the naturalness of the hierarchy of mass
scales in fundamental physics \cite{hierarchy}. It is also known to facilitate the
construction of simple Grand Unified Theories (GUTs) with no intermediate
mass scale, if supersymmetry appears around the TeV scale \cite{gut}. These two 
motivations for low-energy supersymmetry arise specifically in theories with large GUT and 
Planck mass scales, and are supplemented by other
motivations for low-energy supersymmetry, such as cold dark matter \cite{EHNOS} and the
existence of a light Higgs boson \cite{erz}.

Supersymmetry is all very nice, but it must be broken, and there is no
consensus how this occurs. Presumably the origin of
supersymmetry breaking is with a gravitino mass in local supersymmetry \cite{sugr2},
but the mechanism for gravitino mass generation is still unclear, as is
the manner whereby this breaking is communicated to the supersymmetric
partners of observable particles \cite{BIM}. It is often supposed that supersymmetry
is initially broken in some Polonyi or hidden sector of the theory \cite{pol,bfs}, and
is then transmitted to the spartners of Standard Model particles by
either gravitational-strength interactions or some high-scale gauge
interactions.

In phenomenological treatments of supersymmetry, the effective observable
magnitudes of these supersymmetry-breaking parameters at low scales are
then calculated using the renormalization-group equations (RGEs) of the effective
low-energy theory, which is typically taken to be the minimal
supersymmetric extension of the Standard Model (MSSM) \cite{mssm}. 
One often assumes
that the soft supersymmetry-breaking parameters are universal at some high
input scale, and we term the resulting constrained model the CMSSM~\cite{cmssm,efgosi,cmssmnew,cmssmmap,like}.
However, it should be stressed that not all models of supersymmetry
breaking, e.g., in string theory yield such universal input parameters \cite{dterm}.

There is also the question of what input scale should be used to initialize 
the renormalization-group running of the soft supersymmetry-breaking 
parameters. In most CMSSM studies, this is taken to be the supersymmetric GUT scale
$M_{GUT} \sim 2 \times 10^{16}$~GeV, but 
this assumption may be questioned. In general, it should probably be taken 
as approximately equal to the lowest among the dynamical scales in 
the Polonyi or hidden sector where supersymmetry is originally broken, and 
the scales of the interactions that transmit this breaking to the 
observable MSSM particles. 

One could well imagine scenarios in which the input scale is {\it above}
the GUT scale, e.g., if supersymmetry breaking and its mediation are
characterized by the Planck or the string scale. In this case, the soft
supersymmetry-breaking gaugino masses $m_{1/2}$ would evolve together down to the
GUT scale, where they would still be universal, diverging at lower scales
according to the conventional MSSM RGEs.  On
the other hand, the soft supersymmetry-breaking scalar masses $m_0$ would not in
general be universal at the GUT scale $M_{GUT}$, but would be different for
different GUT multiplets. For example, in conventional SU(5) the scalar
masses of the spartners of the $d_R$ and $\ell_L$ would be identical, but
different from those of the spartners of the $q_L, u_R$ and $e_R$, since
they come from $\boldmath{\bar{5}}$ and $\boldmath{10}$ representations,
respectively. On the other hand, in flipped SU(5) the groupings would be
$u_R, \ell_L$ and $q_L, d_R$, with the $e_R$ different again, whereas only
in SO(10) would all the soft supersymmetry-breaking scalar masses of the
quarks and leptons be universal (but not those of the Higgs bosons). These
would be interesting scenarios to study, but are not the objects of this
paper.

Here we study instead the equally (if not more) plausible case in which universality applies
to the parameters $m_{1/2}$ and $m_0$ 
at some input scale {\it below} the GUT scale. This might occur if the 
scale at which supersymmetry is broken dynamically in some hidden sector is smaller than the $M_{GUT}$, for example due to the v.e.v. of
some condensate that appears at a lower scale. A partial analogue may be the chiral-symmetry breaking quark condensate in QCD, which generates a `soft' effective quark mass that `dissolves' at scales above $\Lambda_{QCD}$. Alternatively, perhaps `hard' supersymmetry breaking in the
hidden sector is communicated to the observable sector by loops of particles weighing less than
$M_{GUT}$, which `dissolve' at high scales. In any such sub-GUT CMSSM scenario, the gaugino
masses would evolve in the same way as the gauge couplings at the leading
(one-loop) level, but from a different
starting point, so that their effective values at low energies would be less
separated than they are in the usual GUT CMSSM scenario. Likewise, the
effective values of the soft supersymmetry-breaking scalar masses at low
energies would also be more similar in a sub-GUT CMSSM than in the usual scenario. 

The renormalization of the gauge couplings would always be the same in
sub-GUT CMSSM scenarios, and the successful
coupling unification of supersymmetric GUTs would
therefore be preserved. However, because the renormalizations of the soft
supersymmetry-breaking parameters would differ in these scenarios, as
we demonstrate and explain, the regions of the $(m_{1/2}, m_0)$ plane
allowed by experiments and cosmology in such a sub-GUT CMSSM scenario may
be very different from those allowed in the usual GUT CMSSM scenario. For
example, the impact of the LEP constraint on the MSSM Higgs $h$ is more marked, 
because the reduced dependence on $m_{1/2}$ of $m_{\tilde t}$ (which largely 
controls $m_h$) implies that only values of $m_{1/2}$ larger than those required
in the GUT CMSSM are allowed in a sub-GUT CMSSM.

However, the most dramatic aspect of a sub-GUT CMSSM scenario may be the
altered form of the constraint imposed by the relic density of supersymmetric cold dark matter. 
We assume that R parity is conserved, so that the lightest supersymmetric particle
(LSP) is stable, and hence should be present in the Universe today as a relic from
the Big Bang. We further assume that the lightest supersymmetric particle (LSP) is the 
neutralino  $\chi$. In
the usual GUT CMSSM scenario, one may distinguish three 
well-separated, generic regions of
the $(m_{1/2}, m_0)$ plane that are allowed by the dark matter constraint 
imposed by WMAP~\cite{WMAP} on the relic $\chi$ density: 
the coannihilation region \cite{stauco}, 
the focus-point region \cite{focus} and the
rapid-annihilation funnel region \cite{efgosi,funnel}. In sub-GUT CMSSM models, these regions
tend to merge in a striking way as the input supersymmetry-breaking
scale is reduced. This behaviour is understandable, stemming from the relations
between different MSSM particle masses. In the coannihilation
region, the neutralino and lighter stau have very similar masses, whereas
in the focus-point region $|\mu| \sim m_W$, and in the funnel region
$m_\chi \sim m_A/2$. Because of the different degrees of renormalization
of the sparticle masses in sub-GUT CMSSM models, the relations between
these masses and the underlying parameters $m_{1/2}$ and $m_0$ change,
causing the three different regions to move and ultimately merge.

\section{Experimental, phenomenological and cosmological constraints
in the CMSSM}

\begin{figure}
\begin{center}
\mbox{\epsfig{file=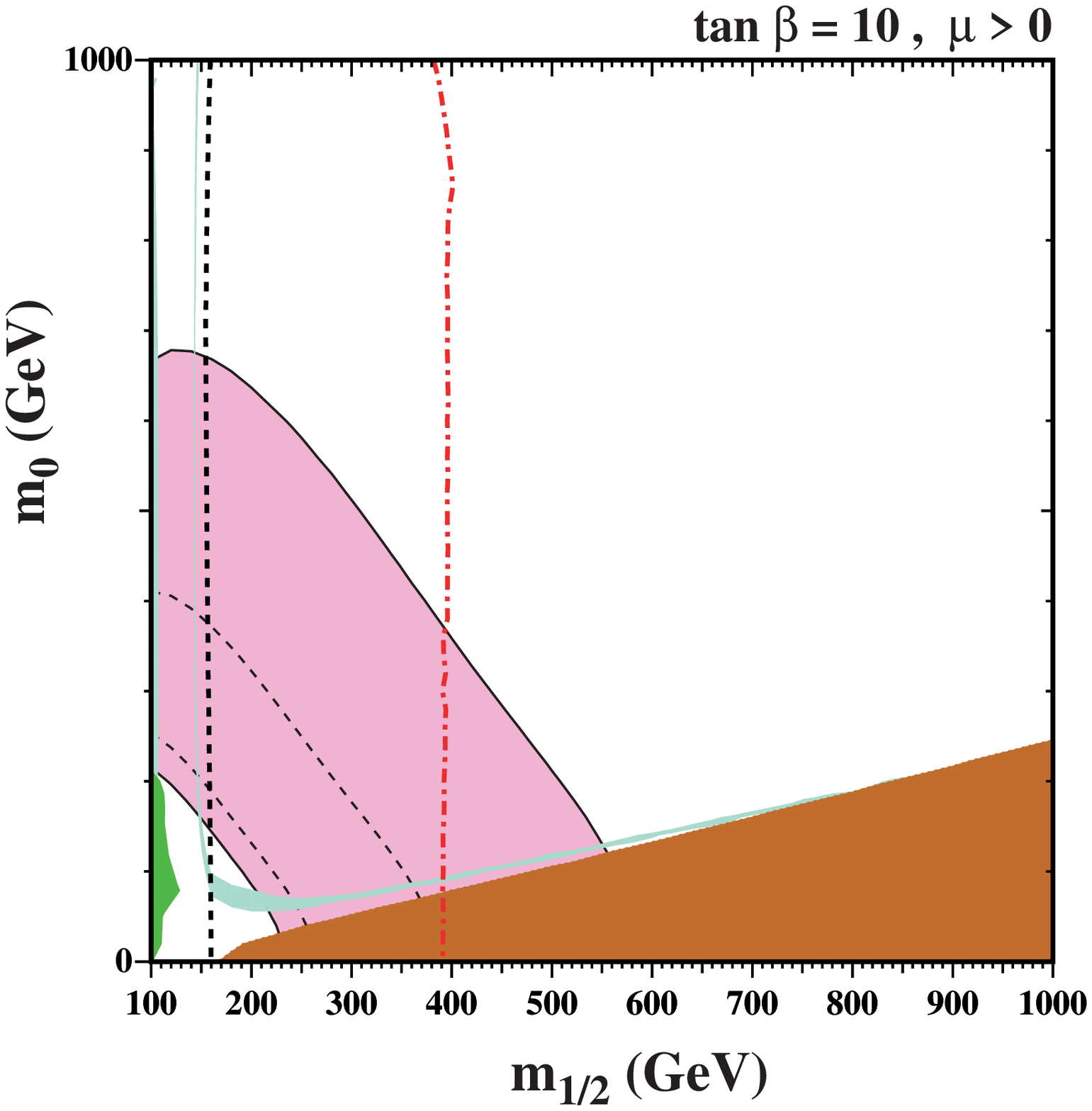,height=7cm}}
\mbox{\epsfig{file=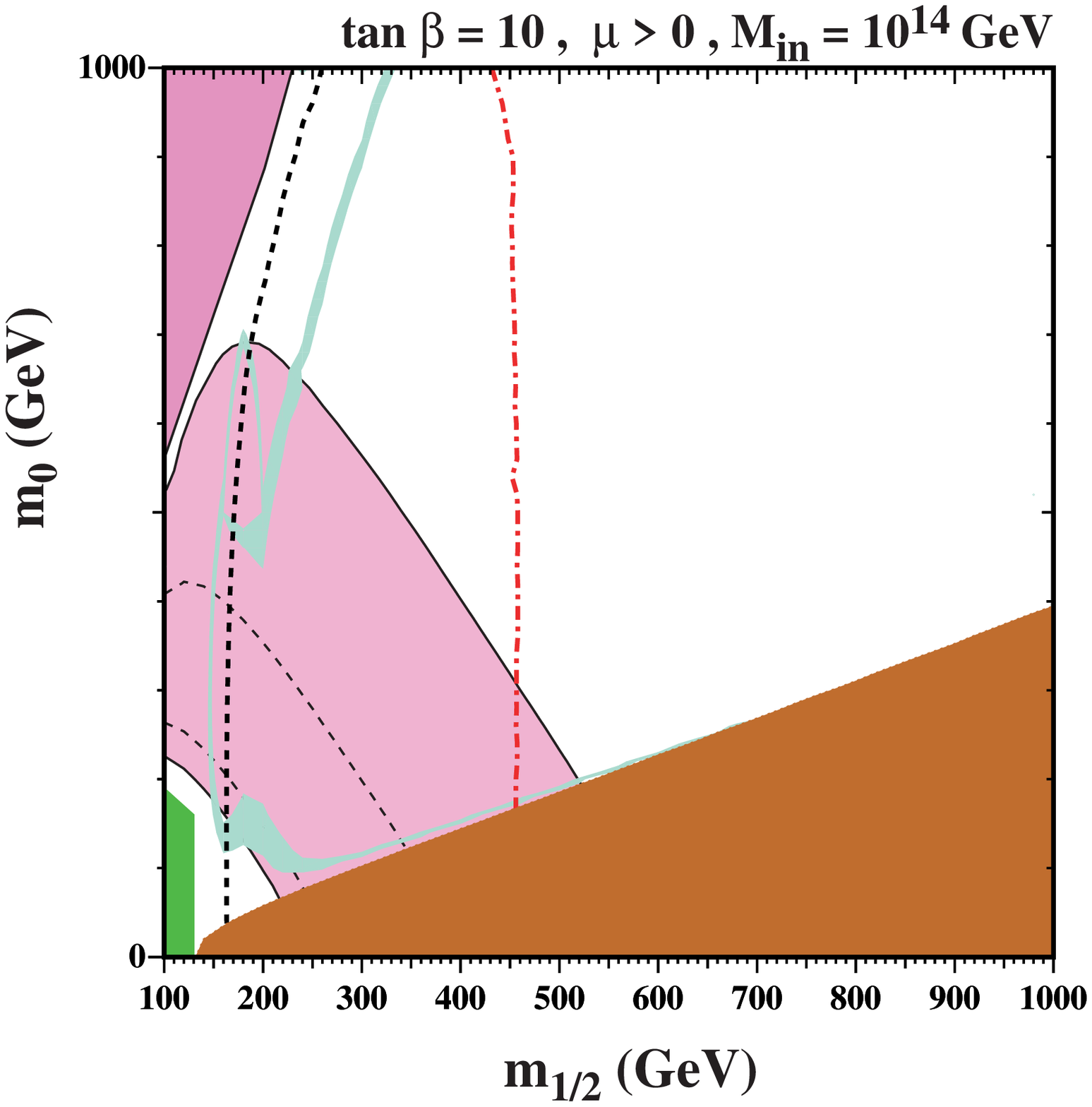,height=7cm}}
\end{center}
\begin{center}
\mbox{\epsfig{file=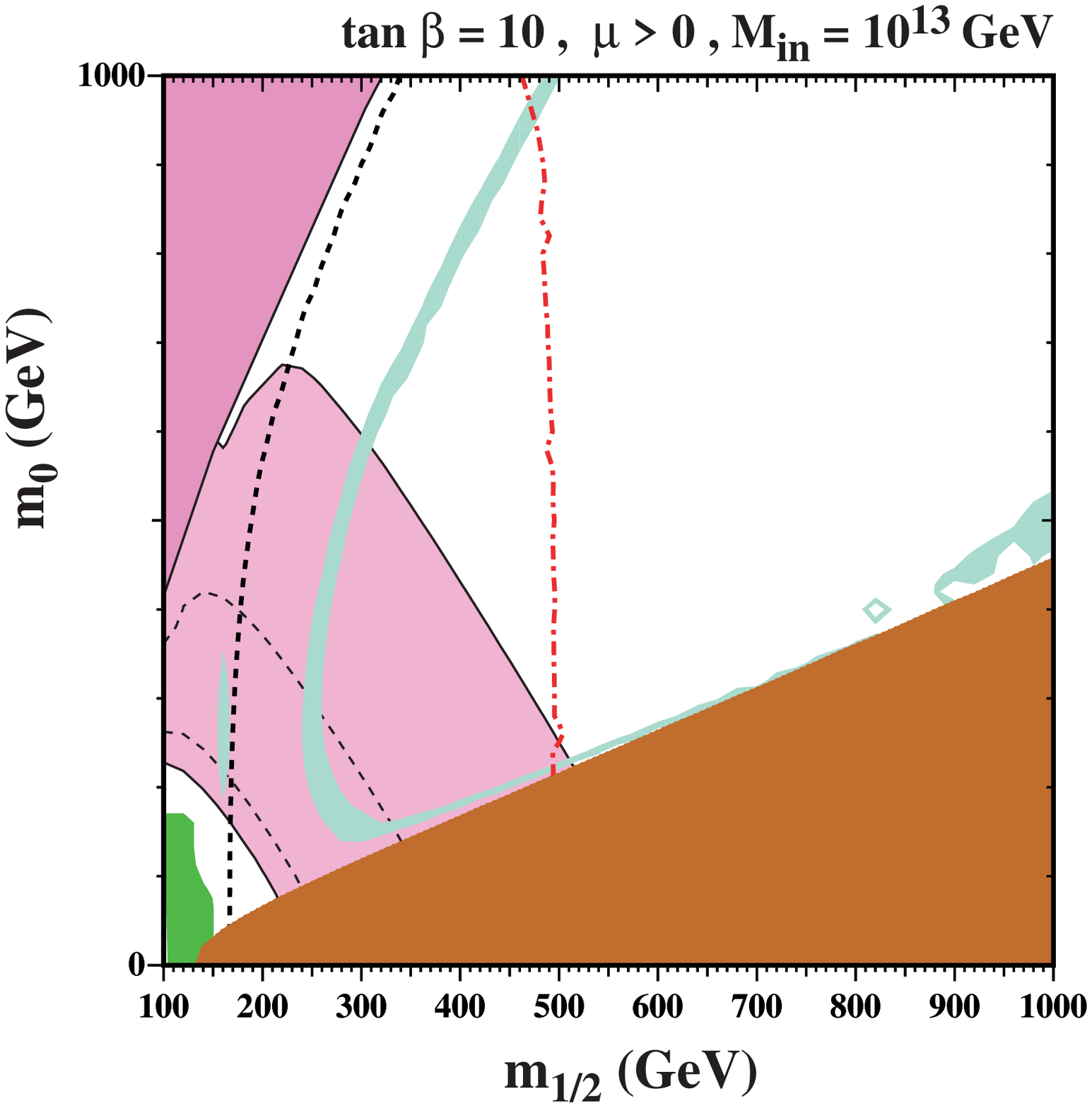,height=7cm}}
\mbox{\epsfig{file=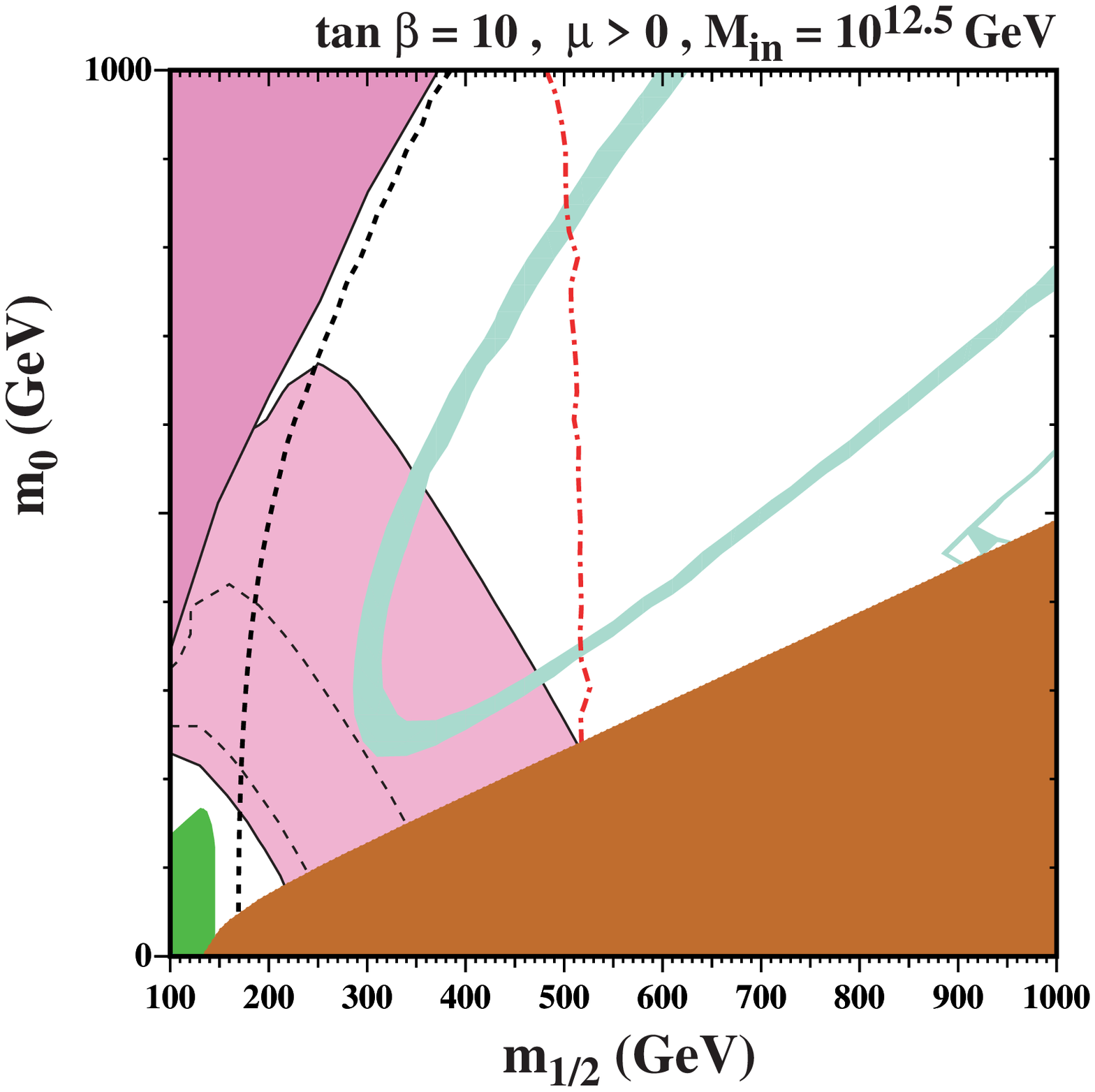,height=7cm}}
\end{center}
\caption{\it
Examples of $(m_{1/2}, m_0)$ planes with $\tan \beta = 10$ and 
$A_0 = 0$ but with  different values of $M_{in}$.
(a) The CMSSM case with $M_{in} = M_{GUT} \sim 2 \times 10^{16}$~GeV, 
(b) $M_{in} = 10^{14}$ GeV, 
(c) $M_{in} = 10^{13}$ GeV and (d) $M_{in} = 10^{12.5}$ GeV. 
In each panel, we show the regions excluded by 
the LEP lower limits on MSSM particles, those ruled out by $b
\to s \gamma$ decay~\protect\cite{bsgex,bsgth} (medium green shading), and those 
excluded 
because the LSP would be charged (dark red shading). The region favoured 
by the WMAP range $\Omega_{CDM} h^2 =
0.1045^{+0.0072}_{-0.0095}$ has light turquoise shading. The region 
suggested by $g_\mu - 2$ is medium (pink) shaded.}
\label{fig:mint}
\end{figure}

We begin by briefly discussing the constraints imposed on a standard 
GUT CMSSM model.  This will serve as a baseline for comparison with
the sub-GUT CMSSM models which are the focus of this paper.
In Fig. \ref{fig:mint}(a), we show the $(m_{1/2}, m_0)$ plane in the
GUT CMSSM model for $\tan \beta = 10$ and $m_t = 172.5$~GeV~\cite{mt}.  
Among the relevant phenomenological constraints shown are  
the limits on the chargino mass: $m_{\chi^\pm} > 104$~GeV~\cite{LEPsusy},
shown as the near-vertical (black) dashed line at low $m_{1/2}$, 
and on the Higgs mass: $m_h >
114$~GeV~\cite{LEPHiggs}, shown as the near-vertical (red) dot-dashed curve
at $m_{1/2} \approx 400$ GeV~\footnote{Here and throughout this paper, we use
{\tt FeynHiggs}~\cite{FeynHiggs} for the calculation of $m_h$. We do not allow for
the possible theoretical and parametric errors in the {\tt FeynHiggs} results, which
would allow values of $m_{1/2} \sim 80$~GeV
smaller for the value of $\tan \beta = 10$ considered here.}.
Another phenomenological constraint is the requirement that
the branching ratio for $b \rightarrow
s \gamma$ be consistent with the experimental measurements~\cite{bsgex}. 
These measurements agree with the Standard Model, and
therefore provide bounds on MSSM particles~\cite{bsgth}  and hence
the $(m_{1/2}, m_0)$ parameter space. 
At $\tan \beta = 10$ and $\mu > 0$, the bound due to $b \rightarrow
s \gamma$ is weak, as is shown by the green shaded region 
at low $m_{1/2}$ and $m_0$.  Typically, the $b\rightarrow s\gamma$
constraint is more important for $\mu < 0$, but it is also relevant for
$\mu > 0$,  particularly when $\tan\beta$ is large. 
Finally, we display with pink shading the 
regions of the $(m_{1/2}, m_0)$ plane that are favoured by
the BNL measurement~\cite{g-2} of $g_\mu - 2$ at the 2-$\sigma$ level, as calculated
in the Standard Model using $e^+ e^-$ data~\footnote{The $\pm 1-\sigma$ range of the
possible supersymmetric contribution to $g_\mu - 2$ is indicated by dashed lines. 
In view of the uncertainty surrounding the Standard Model contribution to $g_\mu - 2$, we
consider the implementation of this constraint as purely indicative.}.

As already mentioned, we assume that R parity is conserved, so that the
LSP is stable, and we further assume that the LSP
is the lightest neutralino $\chi$.
Also shown as the turquoise shaded regions in Fig. \ref{fig:mint} are the parts of
the $(m_{1/2}, m_0)$ plane where the relic density  of the neutralino LSP $\chi$ falls
within the range preferred by  WMAP, 
%and other astrophysical and cosmological data, ! we use wmap alone now.
namely  $0.085 < \Omega_{CDM} < 0.119$ at the 2-$\sigma$
level~\cite{WMAP}.  The cosmological region shown in panel a) corresponds
to the $\chi-{\tilde \tau}$ co-annihilation strip~\cite{stauco}.
The `bulk' region which existed formerly at small $m_{1/2}$
and $m_0$ is excluded for $\tan \beta =10$ with $m_t = 172.5$~GeV
by the Higgs mass bound. 

There is an additional region of acceptable relic 
density in the GUT CMSSM model, known as the
focus-point region~\cite{focus}, which is found
at rather higher values of $m_0$.
As $m_0$ is increased, the value of $\mu$ at the electroweak scale
which is required in the GUT CMSSM to obey
the electroweak symmetry breaking conditions eventually begins to drop. 
When $\mu \la m_{1/2}$, the composition of the LSP gains a strong Higgsino
component, and the relic density begins to drop precipitously.  
As $m_0$ is increased further, there is no longer any 
consistent solution for $\mu$. The focus-point region is not seen in panel a),
since it occurs at $m_0 > 1000$~GeV for the value $m_t = 172.5$~GeV
assumed here. However, the focus-point region does appear
in the sub-GUT CMSSM models discussed below.

Finally, another region of interest is that created by
rapid annihilation via the direct-channel pole mediated by the 
Higgs pseudoscalar  $A$ when $m_\chi
\sim {1\over 2} m_{A}$~\cite{efgosi,funnel}. We recall that the heavier
neutral scalar Higgs boson $H$ is almost degenerate with the pseudoscalar
boson $A$, but plays a much less significant role in the annihilation process.
Since the heavy scalar and pseudoscalar Higgs masses decrease as  
$\tan \beta$ increases, whilst $m_\chi$ is almost fixed by the value of $m_{1/2}$
and is largely independent of $m_0$,
eventually  $ 2 m_\chi \simeq  m_A$ at any fixed value of $m_{1/2}$. The direct-channel
annihilation then becomes rapid, yielding a
`funnel' of parameters with acceptable relic density, that extends to large
$m_{1/2}$ and $m_0$ at large $\tan\beta$.  This region is not 
present in the GUT CMSSM model at $\tan \beta = 10$, but we will see that 
it appears when the input scale for supersymmetry breaking is reduced.
The funnel due to rapid annihilation via the light Higgs scalar is excluded
in this case by the chargino mass bound, as well as by the Higgs mass bound. 

\section{Lowering the universality scale for soft supersymmetry breaking}

We now explore the consequences of reducing below $M_{GUT}$ the scale at which 
universality is assumed for the supersymmetry-breaking parameters $m_{1/2}$ and
$m_0$, as might occur if the underlying supersymmetry-breaking mechanism and/or
the mechanism for communicating it to the observable sector are characterized by
a dynamical scale $M_{in} < M_{GUT}$. One could, in principle, imagine that the 
scales at which the $m_{1/2}$ and $m_0$ parameters are universal
might be different, but we do not consider
such a possibility here~\footnote{We note, in passing, that we also assume universality 
at the same input scale for
the soft trilinear supersymmetry-breaking parameters $A$, though this is not
of great relevance for our discussion.}.

As already mentioned, at the one-loop level the renormalizations of the gaugino
masses $M_a (a = 1, 2, 3)$ are identical with those of the 
corresponding gauge coupling strengths $\alpha_a$, so that in a sub-GUT CMSSM
\begin{equation}
M_a (Q) \; = \; \frac{\alpha_a(Q)}{\alpha_a(M_{in})}M_a(M_{in}),
\label{univino}
\end{equation}
where the input gaugino masses $M_a(M_{in}) =  m_{1/2}$ by
assumption. By comparison, in the usual GUT CMSSM, the values of the gaugino
masses would already be different at the lower scale $M_{in}$:
$M_a(M_{in}) = (\alpha_a(M_{in})/\alpha(GUT)) \times m_{1/2}$. Therefore,
in the sub-GUT CMSSM scenario, the low-energy effective 
soft supersymmetry-breaking gaugino masses differ from each other by smaller amounts
than in the usual GUT CMSSM.

The soft supersymmetry-breaking scalar masses of the different squark and slepton 
flavours and Higgs bosons $m_{0_i}$ are renormalized below the universality scale 
by both gauge interactions and Yukawa interactions. The latter are important
for the stop squarks and the Higgs multiplet 
coupled to them, and for the sbottom squarks, stau sleptons and the other Higgs multiplet
at large $\tan \beta$. The net effects of these renormalizations may be summarized as
follows:
\begin{equation}
m^2_{0_i}(Q) \; = \; m^2_0(M_{in}) + C_i(Q, M_{in}) m^2_{1/2},
\label{suniv}
\end{equation}
where the calculable renormalization coefficients $C_i(Q, M_{in}) \to 0$ as $Q \to M_{in}$,
and, for $M_{in} \ge 10^{11}$~GeV as explored here, $C_i(Q, M_{in}) \to 
C_i(Q, M_{GUT})$ monotonically as $M_{in} \to M_{GUT}$. The coefficients
$C_i(Q, M_{in})$ are positive for all the squarks and sleptons, but negative for the Higgs 
multiplet $H_2$ that is
coupled to the top quark, and also for the other Higgs multiplet $H_1$ at large $\tan \beta$
when it has large couplings to the bottom quark and $\tau$ lepton.
These negative corrections make possible dynamical electroweak symmetry breaking, 
if they drive the full quantity (\ref{suniv}) for the corresponding Higgs multiplet negative at 
low energies. In our
treatment of the sub-GUT CMSSM, we include these effects consistently in the
electroweak vacuum conditions.

\begin{figure}
\begin{center}
\mbox{\epsfig{file=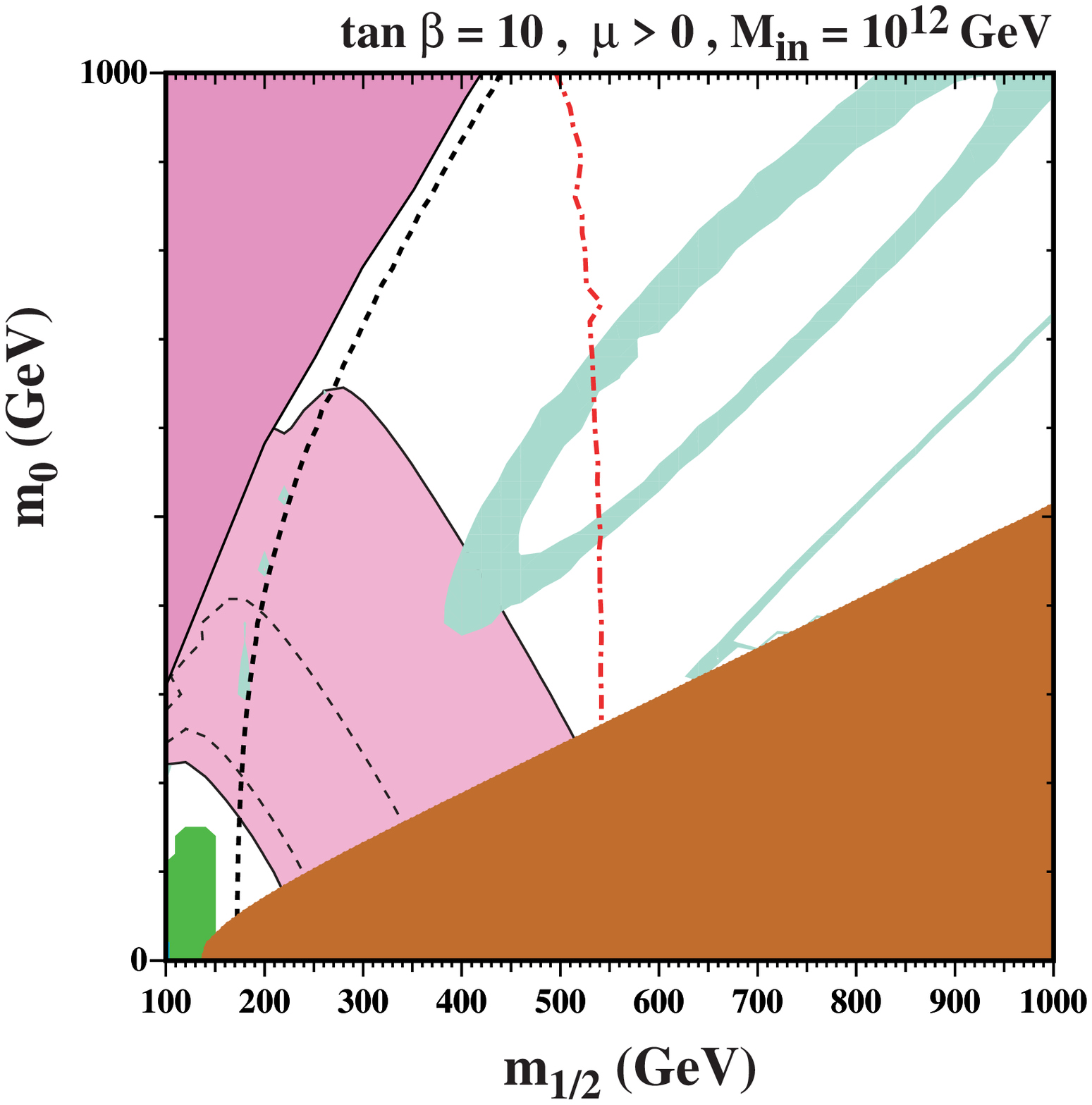,height=7cm}}
\mbox{\epsfig{file=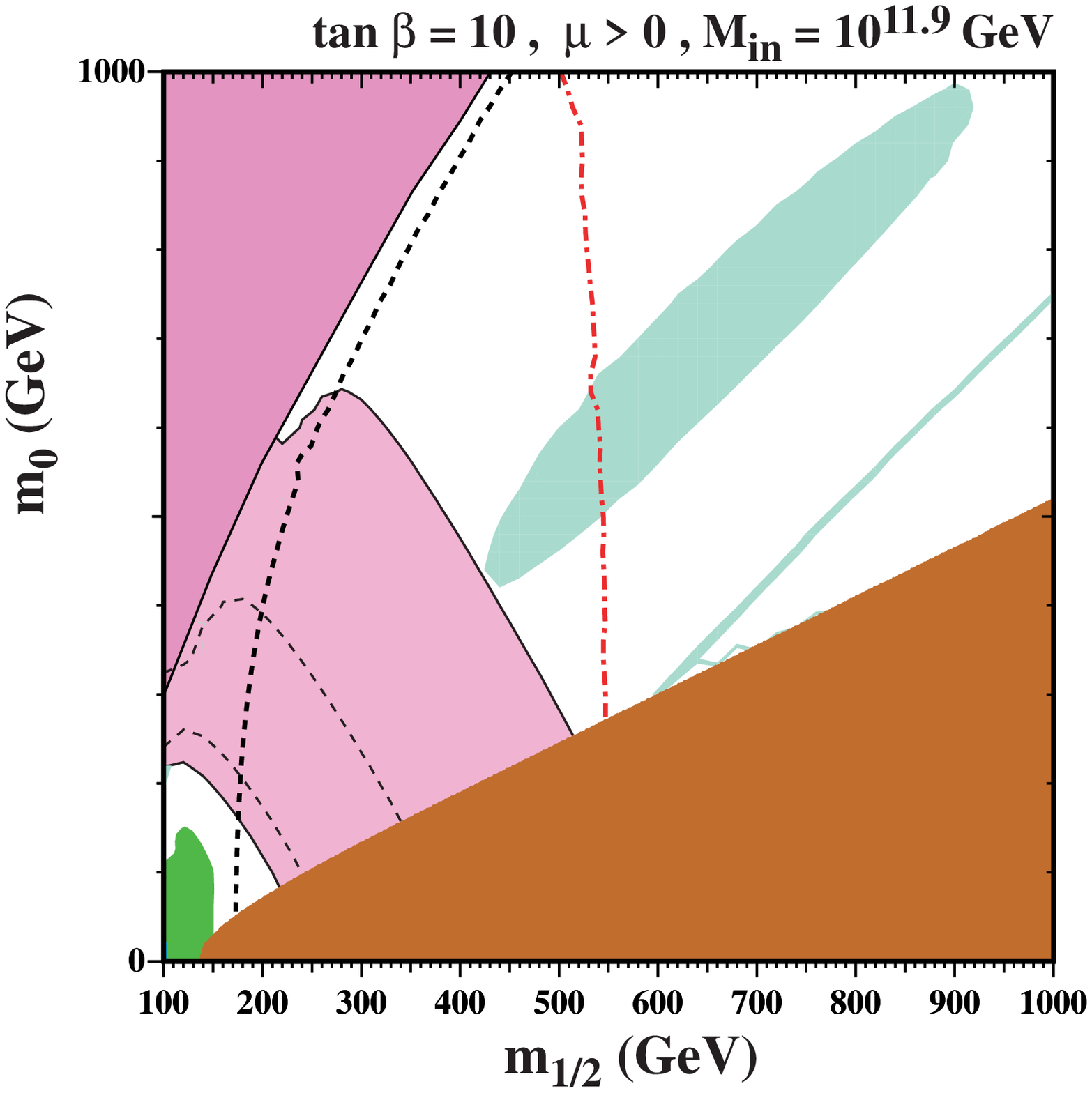,height=7cm}}
\end{center}
\begin{center}
\mbox{\epsfig{file=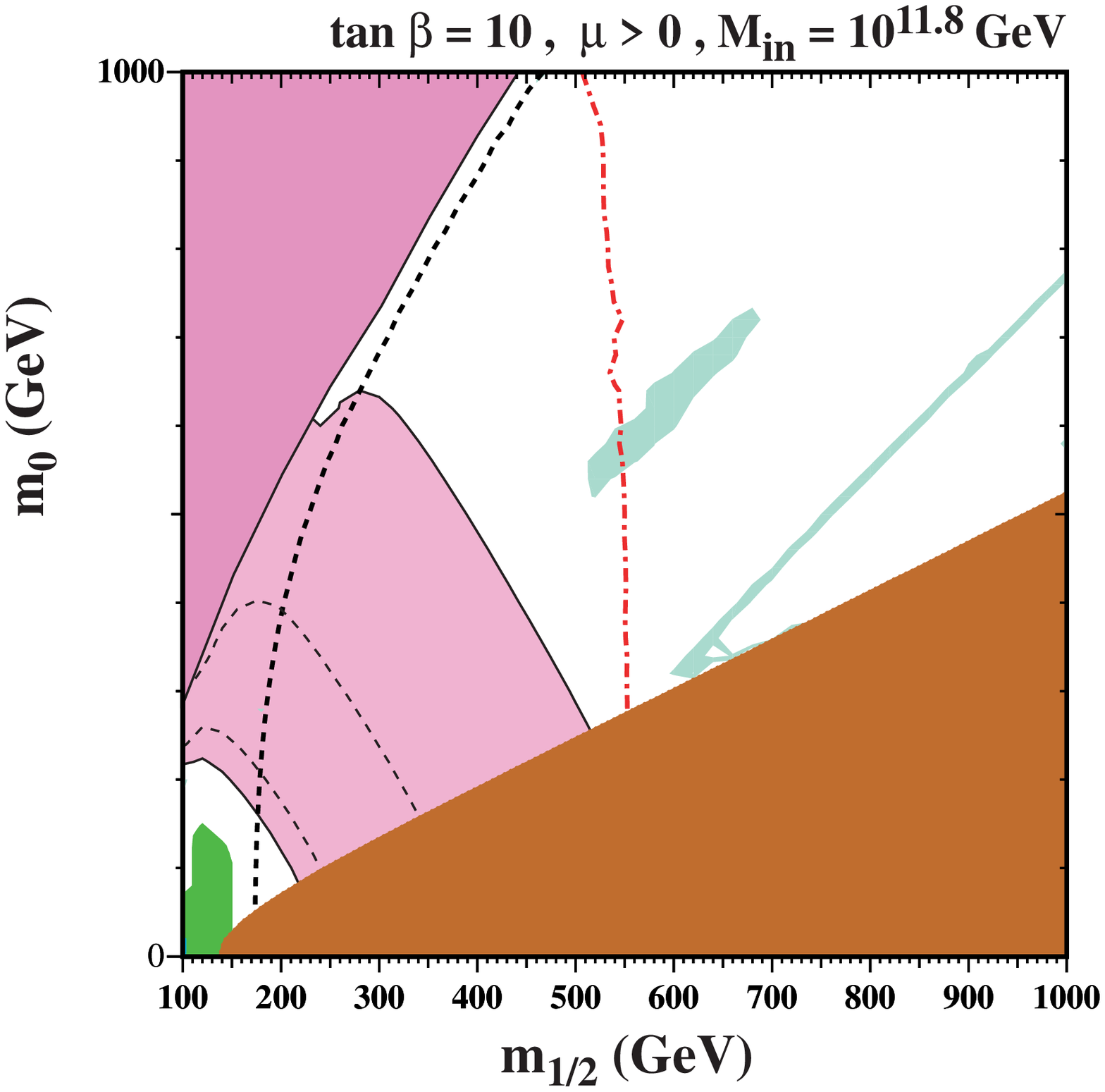,height=7cm}}
\mbox{\epsfig{file=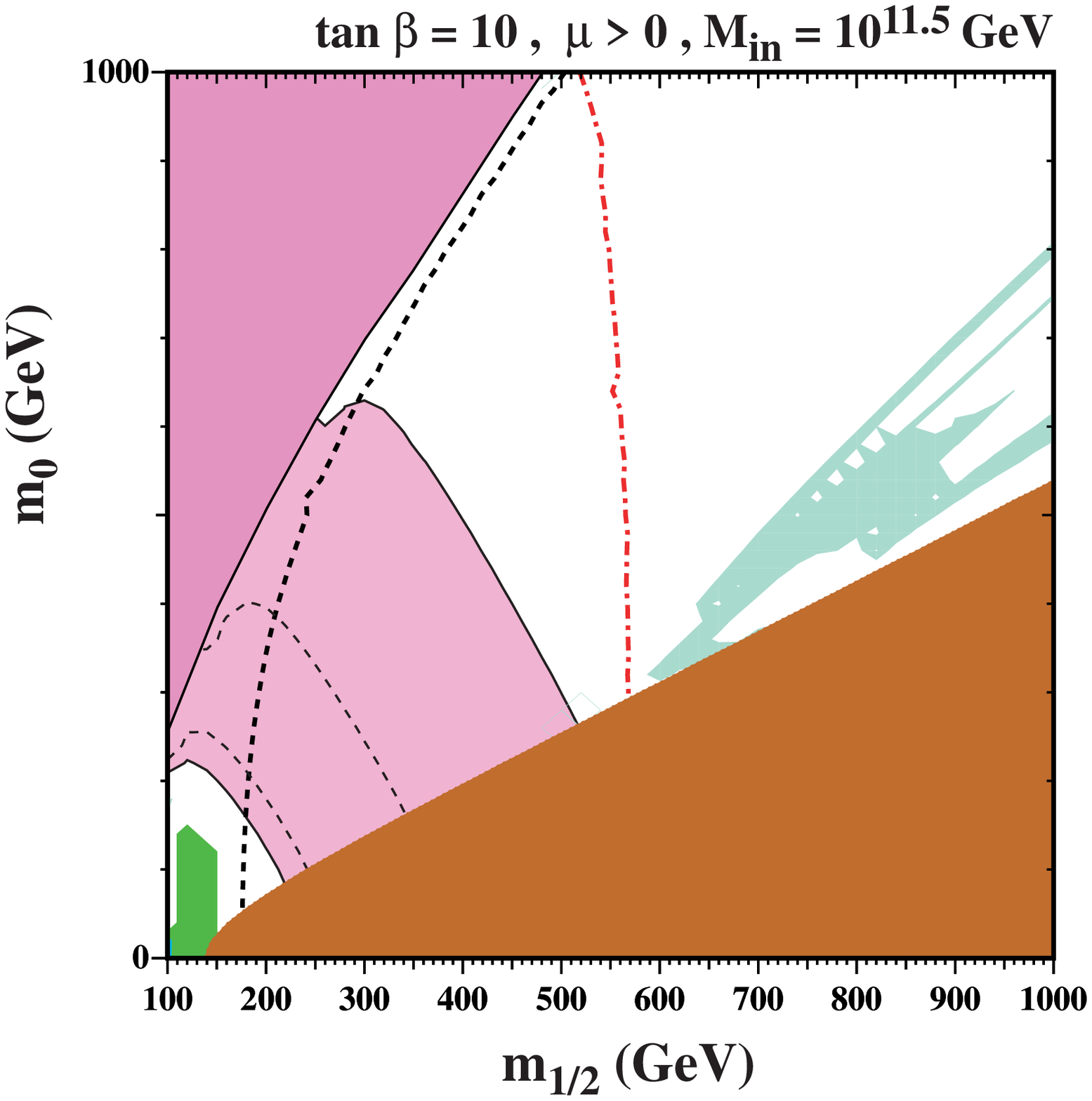,height=7cm}}
\end{center}
\caption{\it
Examples of $(m_{1/2}, m_0)$ planes with $\tan \beta = 10$ and 
$A_0 = 0$ but with  different values of $M_{in}$
(a) $M_{in} = 10^{12}$ GeV , (b) $M_{in} = 10^{11.9}$ GeV, 
(c) $M_{in} = 10^{11.8}$ GeV and (d) $M_{in} = 10^{11.5}$ GeV. 
In each panel, we show the regions excluded by 
the LEP lower limits on MSSM particles, those ruled out by $b
\to s \gamma$ decay~\protect\cite{bsgex,bsgth} (medium green shading), and those 
excluded 
because the LSP would be charged (dark red shading). The region favoured 
by the WMAP range $\Omega_{CDM} h^2 =
0.1045^{+0.0072}_{-0.0095}$ has light turquoise shading. The region 
suggested by $g_\mu - 2$ is medium (pink) shaded.}
\label{fig:mint2}
\end{figure}

We see in Figs.~\ref{fig:mint} and \ref{fig:mint2} several features related to these
renormalization effects. For example, as $M_{in}$ decreases, we see that the requirement
that the LSP not be charged (shown as a brick-red shaded region), 
which imposes the bound $m_{\tilde \tau_1} > m_\chi$
(where ${\tilde \tau_1}$ is the lighter stau slepton),
encroaches on the allowed region of the $(m_{1/2}, m_0)$ plane from the bottom-right
corner. This can be understood from the RGE evolution.  As $M_{in}$ decreases,
the ratio of the lightest neutralino mass to $m_{1/2}$ increases. Simultaneously, 
the coefficient $C_{\tilde {\tau_1}}$ decreases as $M_{in}$ decreases.
Both effects go in the same direction of requiring a higher value of $m_0$
for a given value of $m_{1/2}$ in order to enforce $m_{\tilde \tau_1} > m_\chi$.  
We also see a (purple shaded) bound
that encroaches on the allowed region of the $(m_{1/2}, m_0)$ plane from the top-left
corner, which is due to the change in the electroweak vacuum conditions. The LEP chargino
mass constraint lies just within this boundary, and further within the allowed region is
a strip where $\Omega_\chi$ falls within the WMAP range~\footnote{We return later to
its detailed morphology and evolution with $M_{in}$.} .
This shift in this bound can also be traced directly to the diminished RGE evolution,
and can be understood qualitatively from the tree-level solution for $\mu$:
\beq
\mu^2 = \frac{(m_1^2 - m_2^2 \tan^2 \beta)}{\tan^2 \beta -1} - \frac{M_Z^2}{2}
\label{mu2}
\eeq
where $m_1$ and $m_2$ are the soft Higgs masses associated with
$H_1$ and $H_2$ and the latter is coupled to the top sector\footnote{Note that our results
are based on full two-loop RGEs and not the simple
explanatory approximations given in eqs. \ref{univino} - \ref{mu2}.}. For low
and moderate values of $\tan \beta$, $m_1^2 > 0$ whilst $m_2^2 < 0$
at the weak scale.
As $M_{in}$ decreases, the running of $m_1$ and $m_2$ is suppressed
and, as a result, the absolute values of both remain closer to $m_0$. 
Thus the value of $\mu$ at the weak scale is decreased for any fixed values
of $m_{1/2}$ and $m_0$, and the line where 
$\mu^2$ changes sign is found at a lower value of $m_0$ for any fixed value
of $m_{1/2}$.  The purple shaded
regions in Figs.~\ref{fig:mint} and \ref{fig:mint2} correspond to regions
for which  $\mu^2 < 0$, which are therefore unphysical.   

Finally, we also see that the lower bound
on $m_{1/2}$ due to the LEP Higgs constraint becomes more stringent as $M_{in}$
decreases. This is because $m_h < m_Z$ at the tree level, with a renormalization
that is dominated by a logarithmic dependence on $m_{\tilde t}$. In turn, we see from
(\ref{suniv}) that $m_{\tilde t}$ increases with $m_{1/2}$, at a rate that is suppressed as
$M_{in}$ is decreased. Thus, one requires a 
progressively higher value of $m_{1/2}$ in order to
push the lightest CMSSM Higgs mass above the LEP lower limit $m_h > 114$~GeV.

\section{Evolution of the dark matter constraint}

We now discuss separately the evolving impact of the WMAP 
relic-density constraint as $M_{in}$
is decreased for fixed $\tan \beta = 10$ and $\mu > 0$. 
We see in the usual GUT CMSSM scenario in Fig.~\ref{fig:mint} the familiar
feature of the $\chi - {\tilde \tau_1}$ coannihilation strip at low $m_0$, which extends from
$m_{1/2} \sim 400$~GeV (where it is cut off by the $m_h$ constraint) up to $m_{1/2} \sim
900$~GeV, where it drops down into the forbidden ${\tilde \tau_1}$ LSP 
region~\footnote{If the gravitino were light, the ${\tilde \tau_1}$ would become the NLSP
in this region, and there would be an allowed region with gravitino dark matter, but we do
not explore this possibility here.}. There is no funnel region for this value of $\tan \beta$,
and the focus-point region is unseen at larger values of $m_0$. At low $m_{1/2} \sim
150$~GeV, there is a strip where rapid annihilation via the $h$ pole would bring the
$\chi$ density into the WMAP range which is, however, forbidden by the LEP chargino
constraint and {\it a fortiori} the LEP Higgs constraint.

The picture starts changing already for $M_{in} = 10^{14}$~GeV, as seen in 
Fig.~\ref{fig:mint}(b). The electroweak vacuum condition is visible at $(m_{1/2}, m_0)
\sim (200, 1000)$~GeV, with the chargino constraint close by, and a WMAP strip
tracking its boundary with $m_0 \sim 200$~GeV lower. This WMAP strip does not join
directly with the coannihilation strip, but is instead deflected via a section of the rapid
$h$ annihilation strip at $m_{1/2} \sim 150$~GeV. This behaviour is linked to the
$\chi \chi \to WW$ channel, which has a significant threshold in $m_{1/2}$, but 
whose importance varies with $m_{1/2}$ and $m_0$.  
The rate of variation of the relic density in this
region is reflected in the thickness of the WMAP-allowed region. 
For example, if we follow the relic density at fixed $m_0 = 600$ GeV, we find
that at small $m_{1/2}$, the relic density is low due to the 
rapid annihilation through the light Higgs.  As $m_{1/2}$ is increased,
the density increases and at $m_{1/2} \simeq 170 - 190$ GeV, the density is too high.
However at slightly higher $m_{1/2}$, the $WW$ channel opens up,
and because $\mu$ is lower relative to its value in the GUT-CMSSM, the
relic density drops and becomes small at $m_{1/2} \la 200$ GeV.  
As one moves away from the forbidden triangle in the upper left, $\mu$ begins to 
increase, and the relic density again begins to increase so that 
the relic density is too large when $m_{1/2} \ga 240$ GeV.  Thus, along this horizontal line, we have passed through
three regions for which we match the WMAP relic density.
The coannihilation strip
is rather similar to that in the GUT CMSSM case shown in Fig.~\ref{fig:mint}(a).

There is a more dramatic change for $M_{in} = 10^{13}$~GeV, as seen in Fig.~\ref{fig:mint}(c).
Not only has the electroweak vacuum constraint encroached further on the
$(m_{1/2}, m_0)$ plane, but also the focus-point WMAP strip has receded further away from
it, appearing at $m_0 \sim 300$~GeV lower. Moreover, this focus-point strip now connects 
smoothly at $m_{1/2} \sim 250$~GeV with the $\chi - {\tilde \tau_1}$ coannihilation strip at 
low $m_0$. The coannihilation strip itself exhibits some broadening and embryonic 
bifurcation at $m_{1/2} \sim 1000$~GeV, due to the approaching funnel.

The emerging picture is much clearer in Fig.~\ref{fig:mint}(d), where $M_{in} = 
10^{12.5}$~GeV. The focus-point part of the 
WMAP strip has now separated further from the electroweak vacuum boundary, but also
the linked `coannihilation' portion of the WMAP strip has separated from the
${\tilde \tau_1}$ LSP boundary, by an amount that increases with $m_{1/2}$. In fact,
we now recognize the region at large $m_{1/2}$ as the opening of a characteristic
rapid $A, H$ annihilation funnel, of the type seen in the GUT CMSSM only when
$\tan \beta \sim 50$ for $\mu > 0$ as studied here. On the further side of the funnel,
at $m_{1/2} \sim 900$~GeV, we now see more clearly the bifurcation of the second
funnel wall from the continuing coannihilation strip.

The changes described above accelerate as $M_{in}$ decreases further, as seen in
Fig.~\ref{fig:mint2}. For $M_{in} = 10^{12}$~GeV, as seen in Fig.~\ref{fig:mint2}(a),
the former focus-point, lower coannihilation and funnel regions merge into a WMAP
ellipse that encloses just a small region where the $\chi$ relic density is too large.
The further wall of the funnel and the continuation of the coannihilation strip form a
well-developed `vee' shape that extends to much larger values of $m_{1/2}$ than
those shown here.

Even more strikingly, when $M_{in}$ is reduced slightly to $10^{11.9}$~GeV, as
shown in Fig.~\ref{fig:mint2}(b), the ellipse is now filled up. This is the culmination of
a trend, noticeable already in Fig.~\ref{fig:mint}, for the WMAP regions to
broaden as well as merge as $M_{in}$ decreases. The possibility that the LSP
relic density falls within the WMAP range therefore appears more `natural'.
Moreover, we see in Fig.~\ref{fig:mint2}(a), (b) that it is increasingly `unlikely' that
the relic density will exceed the WMAP range, whereas this appeared much
more `likely' in the GUT CMSSM case shown in Fig.~\ref{fig:mint}(a). Whether one
worries about the `naturalness' of supersymmetric dark matter or not, it is
nevertheless interesting that there is less cause for worry when $M_{in}
\sim 10^{12}$~GeV.

The situation changes again with just a small change to $M_{in} = 10^{11.8}$~GeV,
as seen in Fig.~\ref{fig:mint}. The ellipse has now almost evaporated, with the
relic density falling below the range favoured by WMAP over most of the visible part of
the $(m_{1/2}, m_0)$ plane~\footnote{These regions would of course still be acceptable for
cosmology, if there were another important source of cold dark matter.}. The only region
with an excessive amount of cold dark matter is inside the `vee' at large $m_{1/2}$.
Note also, that the region favoured by the relic density no longer overlaps with
the region preferred by  the $g_\mu - 2$ anomaly.   

Finally, when $M_{in} = 10^{11.5}$~GeV, as shown in Fig.~\ref{fig:mint}(d), the
ellipse favoured by WMAP has disappeared completely We also notice that the
large-$m_{1/2}$ `vee' starts to fill in, with a new generic region of acceptable relic
density now appearing. This is due, in particular, to the opening up of new 
annihilation channels such $(H,A) + Z, H^\pm + W^\mp$ that are sufficient to bring 
the relic density down into the WMAP range. At lower values of $M_{in} \to 10^{10}$~GeV
(not shown), the electroweak vacuum boundary continues to press downwards and the relic 
density is always below the favoured WMAP range for $m_\chi < m_A/2$. 
The relic density lies within
the WMAP range only along narrow strips close to the top and bottom of the `vee' where $m_\chi
\ge m_A/2, m_{\tilde \tau_1}$.
To better understand this behaviour, let us look at the density at fixed $m_{1/2} = 900$ GeV.
At large $m_0$, the annihilation cross section is large dominated by the 
broad s-channel pole through the heavy Higgses, $H$ and $A$.  
As $m_0$ is lowered, $2m_\chi$ becomes larger than $m_A$, and at $m_0 \approx 700$,
the WMAP density is attained.  As one moves to lower $m_0$, away from the pole, the relic density 
increases, but the heavy Higgs masses decrease opening up the $H^\pm + W^\mp$
channel when $m_0 \approx 630$ GeV and the $(H,A) + Z$ at slightly lower $m_0$.
In this region of the parameter space, the $s$-wave annihilation cross section is 
dominant and decreases as $m_0$ is lowered, so there is a modest increase in the density 
and the WMAP value is obtained again when $m_0 \la 600$ GeV.
At still lower $m_0$, yet another channel opens up. At $m_0 \la 560$ GeV, the $h,A$ channel
is open and the density once again drops below the WMAP value. 
As we continue to move off of the Higgs funnel, the $h,A$ contribution slowly decreases
and the density rises and surpasses the WMAP value. At this value of $m_{1/2}$, 
we are past the endpoint of $\chi-{\tilde \tau}$ coannihilation and the density is too large as 
we enter the $\tilde \tau$ LSP region. 

If we continue to lower the supersymmetry breaking input scale, $M_{in}$, 
we find that the region seen in Fig.~\ref{fig:mint}(d) begins to evaporate.
At $M_{in} = 10^{11.2}$~GeV, it is gone, but the $\chi-{\tilde \tau}$ coannihilation region
has returned for $M_{1/2} \ga 600$ GeV. The lower end of the coannihilation region continues
to move to higher $M_{1/2}$ as $M_{in}$ is decreased, so that when $M_{in} < 10^{10}$ GeV,
the lower end of the coannihilation region is at $M_{1/2} \approx 900$ GeV.

\section{Evolution of sparticle masses}

We now discuss the extent to which the results presented in the previous Section
can be understood in terms of the evolution of sparticle masses with $M_{in}$,
and the corresponding implications for and of sparticle measurements at colliders such as
the LHC.

\begin{figure}
\begin{center}
\mbox{\epsfig{file=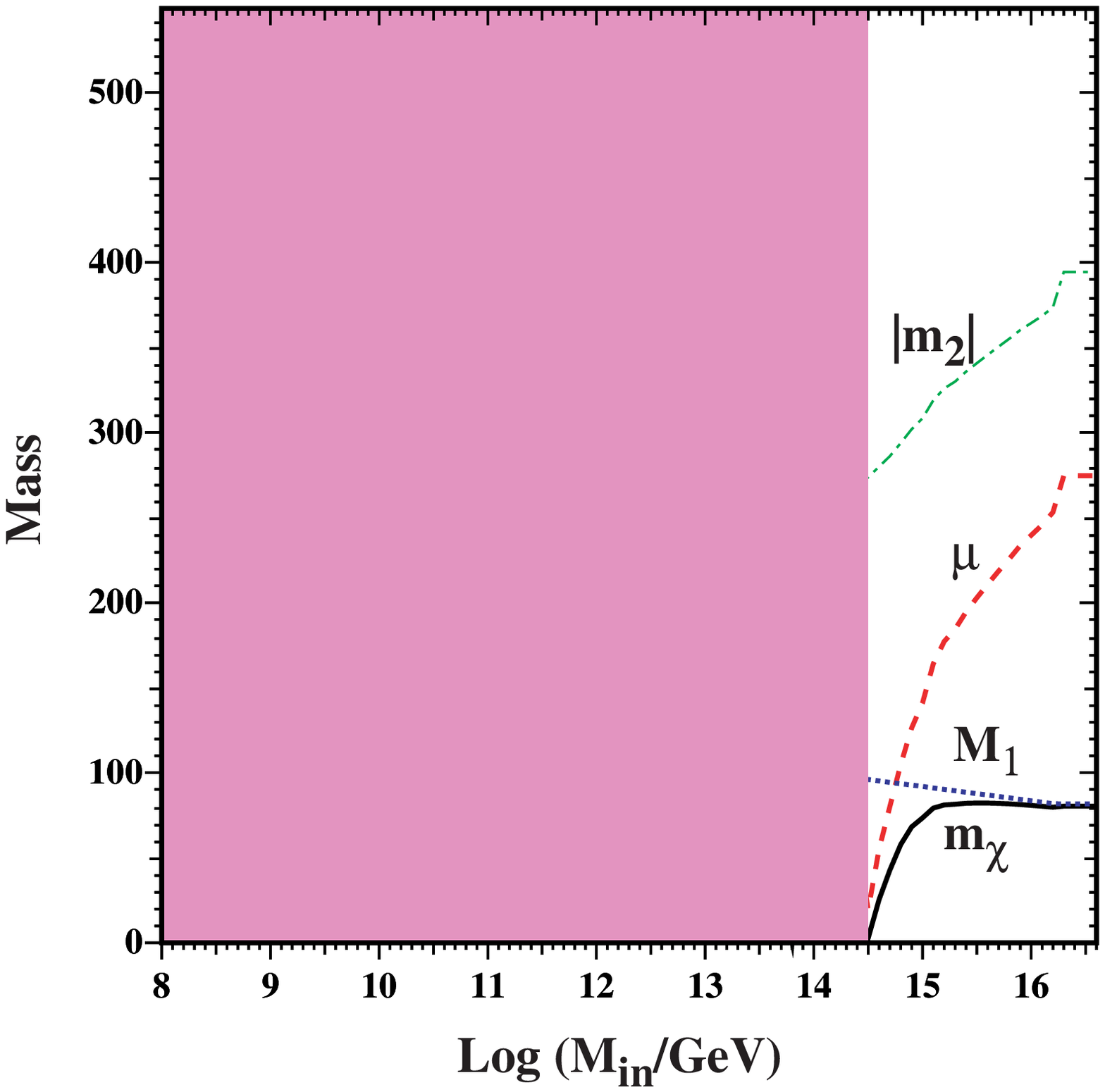,height=7cm}}
\mbox{\epsfig{file=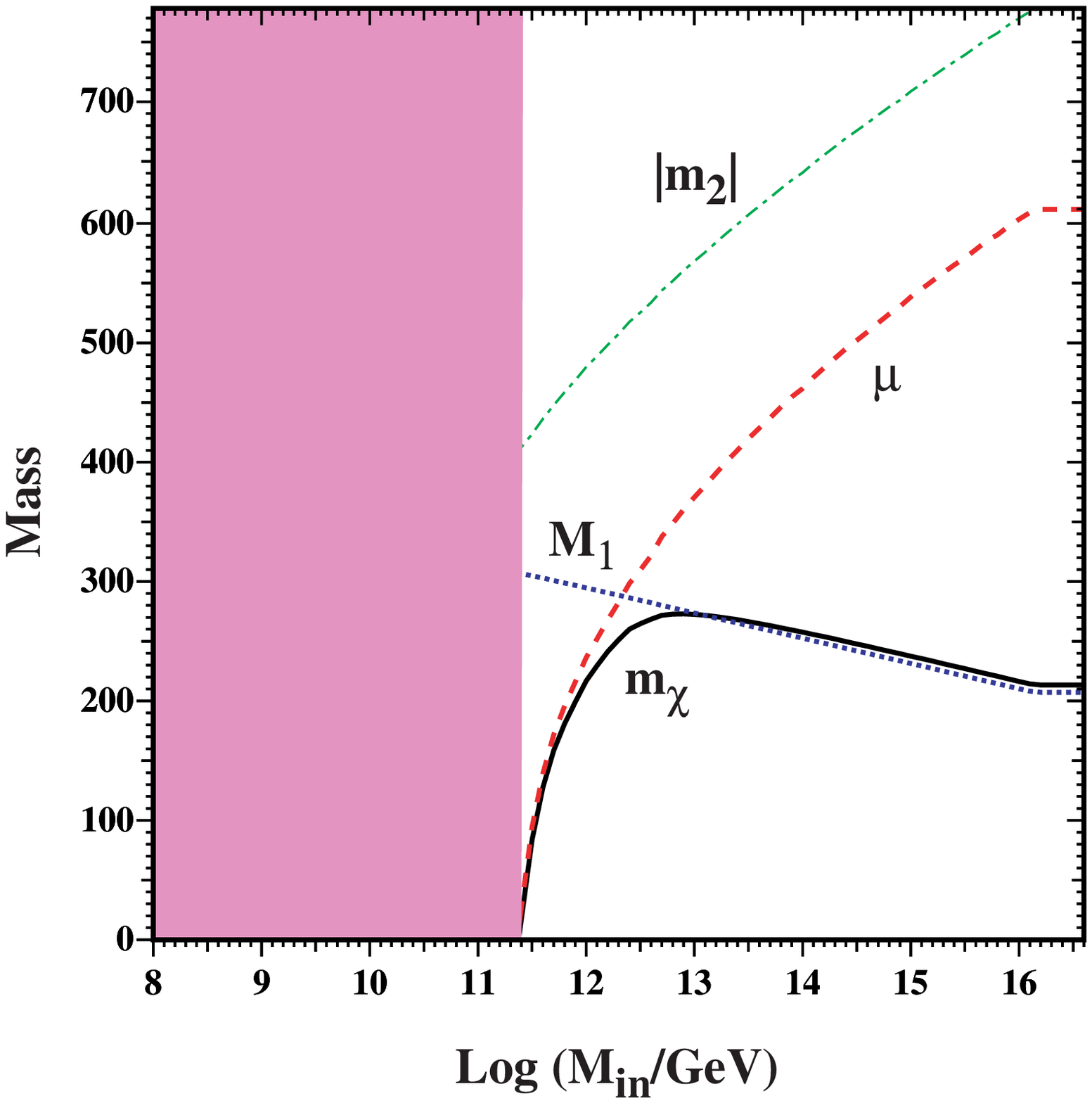,height=7cm}}
\end{center}
\begin{center}
\mbox{\epsfig{file=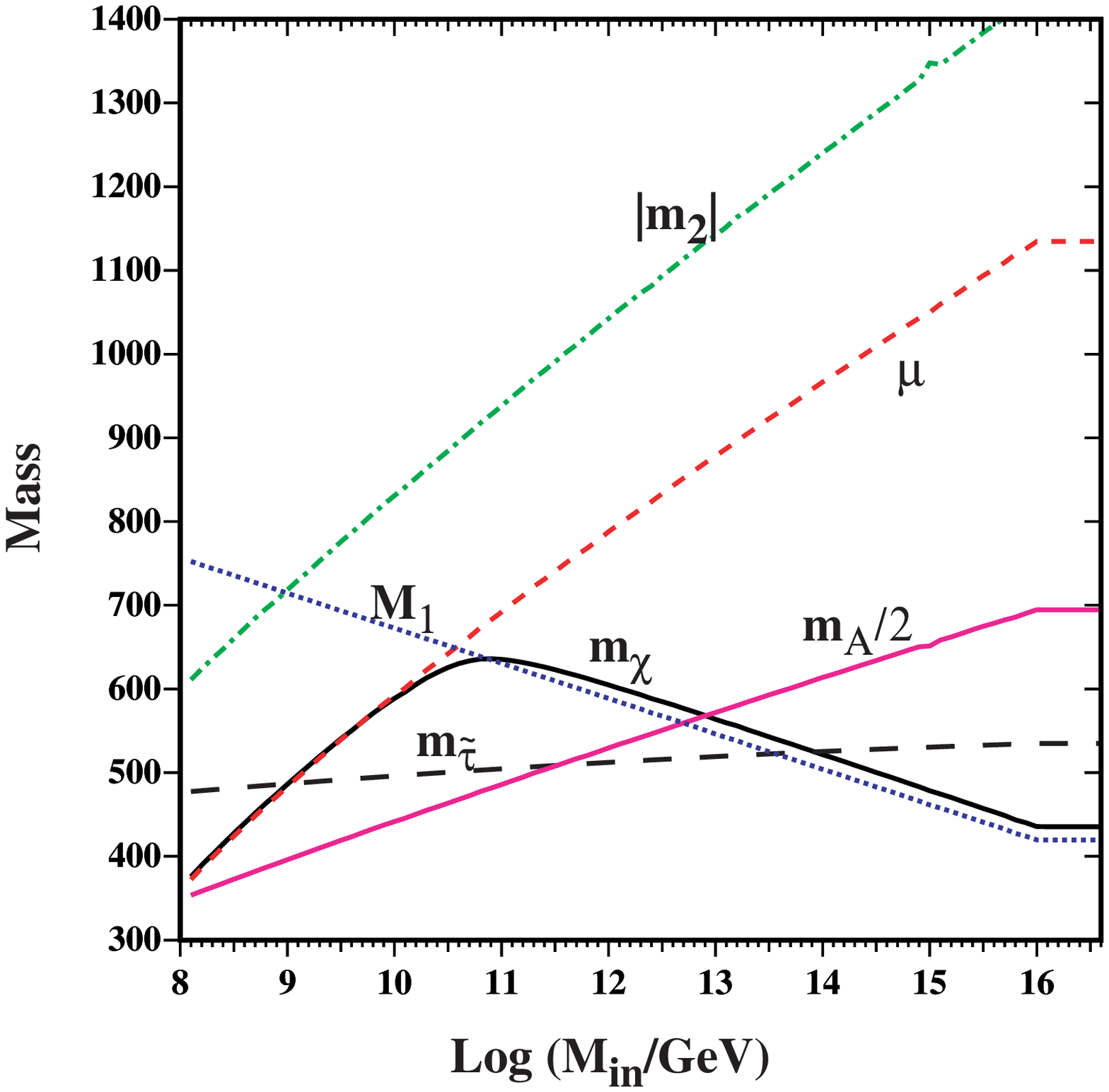,height=7cm}}
\mbox{\epsfig{file=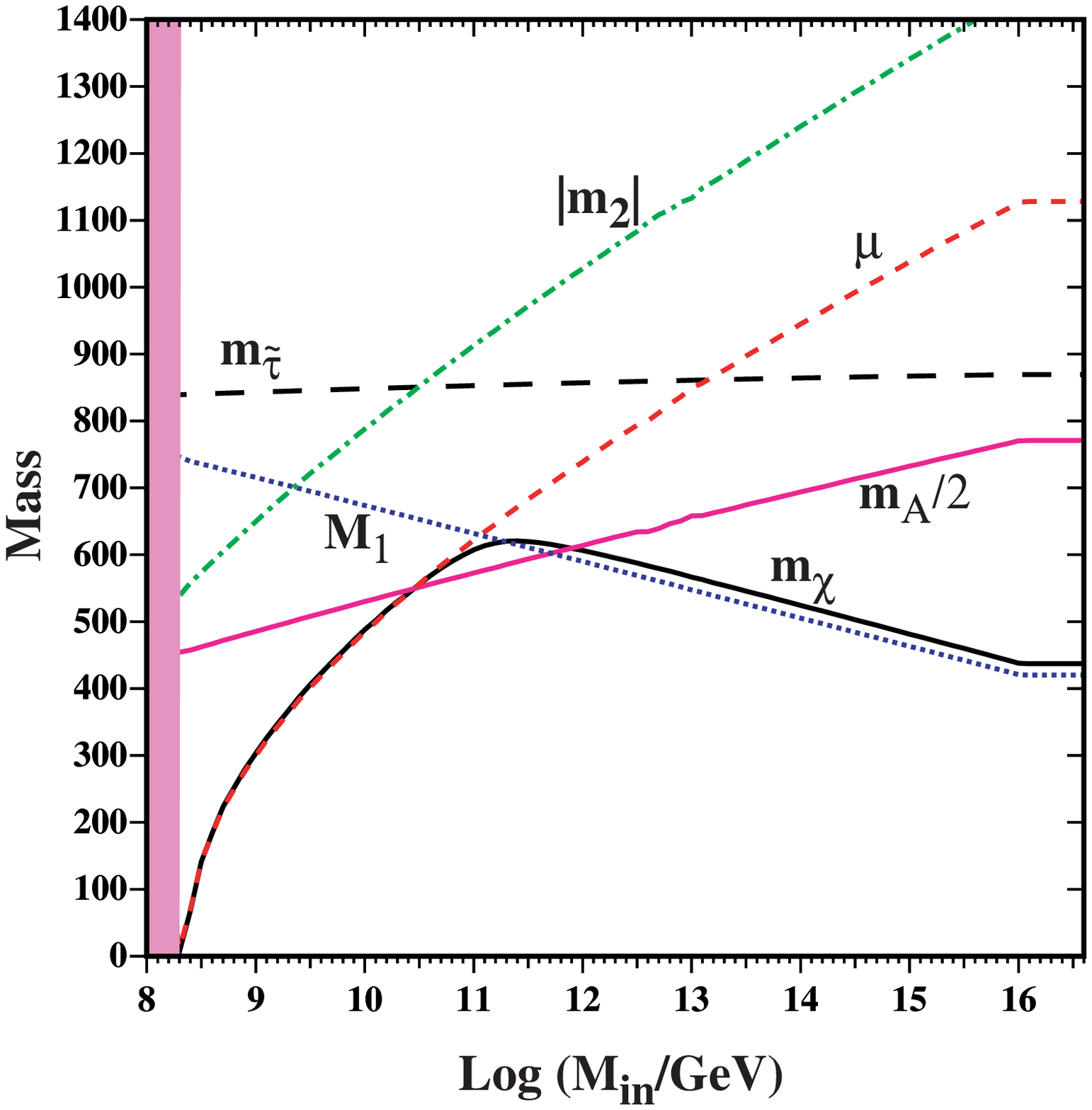,height=7cm}}
\end{center}\caption{\it
Mass parameters as functions of $M_{in}$ in the focus-point
region, for (a) $(m_{1/2}, m_0)
= (200, 1000)$~GeV, and (b) $(m_{1/2}, m_0) = (500, 1000)$~GeV, and in
the funnel region for (c) $(m_{1/2}, m_0)
= (1000, 400)$~GeV, and (d) $(m_{1/2}, m_0) = (1000, 800)$~GeV.}
\label{fig:linear}
\end{figure}

We display in Fig.~\ref{fig:linear} two examples of the evolution of sparticle mass
parameters with $M_{in}$ in the focus-point region. Panel (a) is for $(m_{1/2}, m_0)
= (200, 1000)$~GeV, and panel (b) for $(m_{1/2}, m_0) = (500, 1000)$~GeV. In
each case, we show the evolution of the unmixed electroweak gaugino mass
$M_1$ (blue dotted lines), the Higgs soft mass represented by $sgn(m_2^2)(\sqrt{|m_2^2|})$ 
(turquoise dot-dashed lines), the absolute value of $\mu$ (red dashed lines)
and the LSP mass $m_\chi$ (solid black line). 
We see that, as $M_{in}$ decreases from the GUT
value of $2 \times 10^{16}$~GeV, both $|\sqrt{m_2^2}|$ and particularly $|\mu|$
plummet precipitously, whereas the gaugino masses $M_{1,2}$ evolve more slowly.
In the GUT CMSSM, $m_\chi$ is essentially equal to $M_1$, but this changes as
$M_{in}$ decreases, and $m_\chi$ is given by $|\mu|$ when this is small. In both
the examples shown, the first disaster to occur as $M_{in}$ decreases is that
$|\mu|$ vanishes, which marks the boundary of the region of the $(m_{1/2}, m_0)$
plane allowed by the electroweak vacuum conditions. The disallowed regions are
shaded (purple): this boundary reaches the point $(m_{1/2}, m_0) = (200, 1000)$~GeV 
shown in panel (a) when $M_{in} \sim 10^{14.5}$~GeV, whereas Armageddon
is postponed until $M_{in} \sim 10^{11.4}$~GeV for the point $(m_{1/2}, m_0) = 
(500, 1000)$~GeV shown in panel (b). In both the cases studied, $\sqrt{|m_2^2|}$
does not vanish until well inside the region disallowed by the electroweak vacuum
conditions. We have seen the consequences of this behavior 
in Figs. \ref{fig:mint} and \ref{fig:mint2} as the encroachment of the 
region where the electroweak symmetry breaking conditions are not obeyed.

In panel (a), the relic neutralino density exceeds the WMAP upper
limit in the GUT CMSSM, and the relic density falls as $M_{in}$ decreases. There
is a narrow range of $M_{in} \sim 10^{15}$~GeV where the density falls within
the favoured WMAP range, and it then falls to zero as $|\mu|$ and hence 
$m_\chi$ vanishes. In panel (b), there is a similar sequence of events, with the
WMAP range attained at a lower value of $M_{in} \sim 10^{13.2}$~GeV.

Panels (c) and (d) of Fig.~\ref{fig:linear} provide analogous displays of the
evolution of mass parameters with $M_{in}$ in the funnel region, for
$(m_{1/2}, m_0) = (1000, 400)$~GeV and $(m_{1/2}, m_0) = (1000, 800)$~GeV,
respectively. Here, in addition to $M_1$, $\mu$, $|\sqrt{m_2^2}|$ and $m_\chi$, we also 
plot $m_{\tilde \tau_1}$ and $m_A/2$. The evolution of
$m_{\tilde \tau_1}$ is undramatic. 
As in panels (a) and (b), the physical region is bounded by the vanishing of $\mu$ and hence
$m_\chi$, which occurs at $M_{in} \sim 10^{6.5}$~GeV and $M_{in} \sim 10^{8.3}$~GeV
in cases (c) and (d), respectively. As in the cases (a) and (b), the LSP mass tracks
$M_1$ at large $M_{in}$ and then $\mu$ at smaller $M_{in}$ after the values of
$\mu$ and $M_1$ cross.

Among the more interesting aspects of panels (c) and (d) are the comparisons
between $m_\chi$ and $m_{\tilde \tau_1}$, on the one hand, and between
$m_\chi$ and $m_A/2$, on the other hand. In panel (c), we see that $m_\chi$
rises above $m_{\tilde \tau_1}$ 
(which is unacceptable) when $M_{in}$ falls to $\sim 10^{14}$~GeV, 
a feature visible also in panel (b) of Fig.~\ref{fig:mint}, where we notice that
the point $(m_{1/2}, m_0) = (1000, 400)$~GeV sits on the boundary of the
stau LSP region for this value of $M_{in}$. We also note that $m_\chi$ falls
(with $\mu$) below $m_{\tilde \tau_1}$ when $M_{in} < 10^9$~GeV, an effect not
visible in our previous scans of the $(m_{1/2}, m_0)$ planes
in Figs.~\ref{fig:mint} and \ref{fig:mint2}, where we only
considered $M_{in} \ge 10^{11.5}$~GeV. In panel (d), we again see the
crossover from $m_\chi \sim M_1$ to $m_\chi \sim \mu$, whereas $m_{\tilde \tau_1}
> m_\chi$ in this case.

Comparing now $m_\chi$ with $m_A/2$, we see in panel (c) that in the
case $(m_{1/2}, m_0) = (1000, 400)$~GeV they become equal only in the
stau LSP region when $M_{in}  \sim 10^{13}$~GeV, whereas in the case
$(m_{1/2}, m_0) = (1000, 800)$~GeV shown in panel (d) $m_\chi$ and $m_A/2$
become equal twice, when $M_{in} \sim 10^{12}$ and $10^{10.5}$~GeV, and
$m_\chi$ and $m_A/2$ are quite similar for intermediate and adjacent values
of $M_{in}$.

Since the relation between $m_\chi$ and $m_{\tilde \tau_1}$ is very important
for coannihilation, and that between $m_\chi$ and $m_A/2$ is very important
for the rapid-annihilation funnel, these crossover patterns have important
effects on the relic $\chi$ density, and enable us to understand some
features of Figs.~\ref{fig:mint} and \ref{fig:mint2}. Specifically, for 
$(m_{1/2}, m_0) = (1000, 400)$~GeV as shown in panel (c) of Fig.~\ref{fig:linear},
the approach towards $m_\chi = m_{\tilde \tau_1}$ as $M_{in} \to 10^{14}$~GeV
is responsible for a significant reduction in the dark matter density. The relic
density is also reduced for the case $(m_{1/2}, m_0) = (1000, 800)$~GeV
shown in panel (d) of Fig.~\ref{fig:linear} as $M_{in} \to 10^{12.5}$~GeV,
as also seen in Fig.~\ref{fig:mint}. The relic density then remains below the
range favoured by WMAP as $M_{in} \to 10^{11.8}$~GeV, as seen in the
first three panels of Fig.~\ref{fig:mint2}. On the other hand, the density rises to
the favoured WMAP range when $M_{in} = 10^{11.5}$~GeV, and would even
exceed the WMAP range for smaller values of $M_{in}$. This is because $m_\chi$
is now greater than $m_A/2$. However, we expect the density to fall again as
$M_{in}$ decreases further and $m_\chi$ decreases again and crosses $m_A/2$
a second time.

\section{Implications for collider searches}

\begin{figure}
\begin{center}
\mbox{\epsfig{file=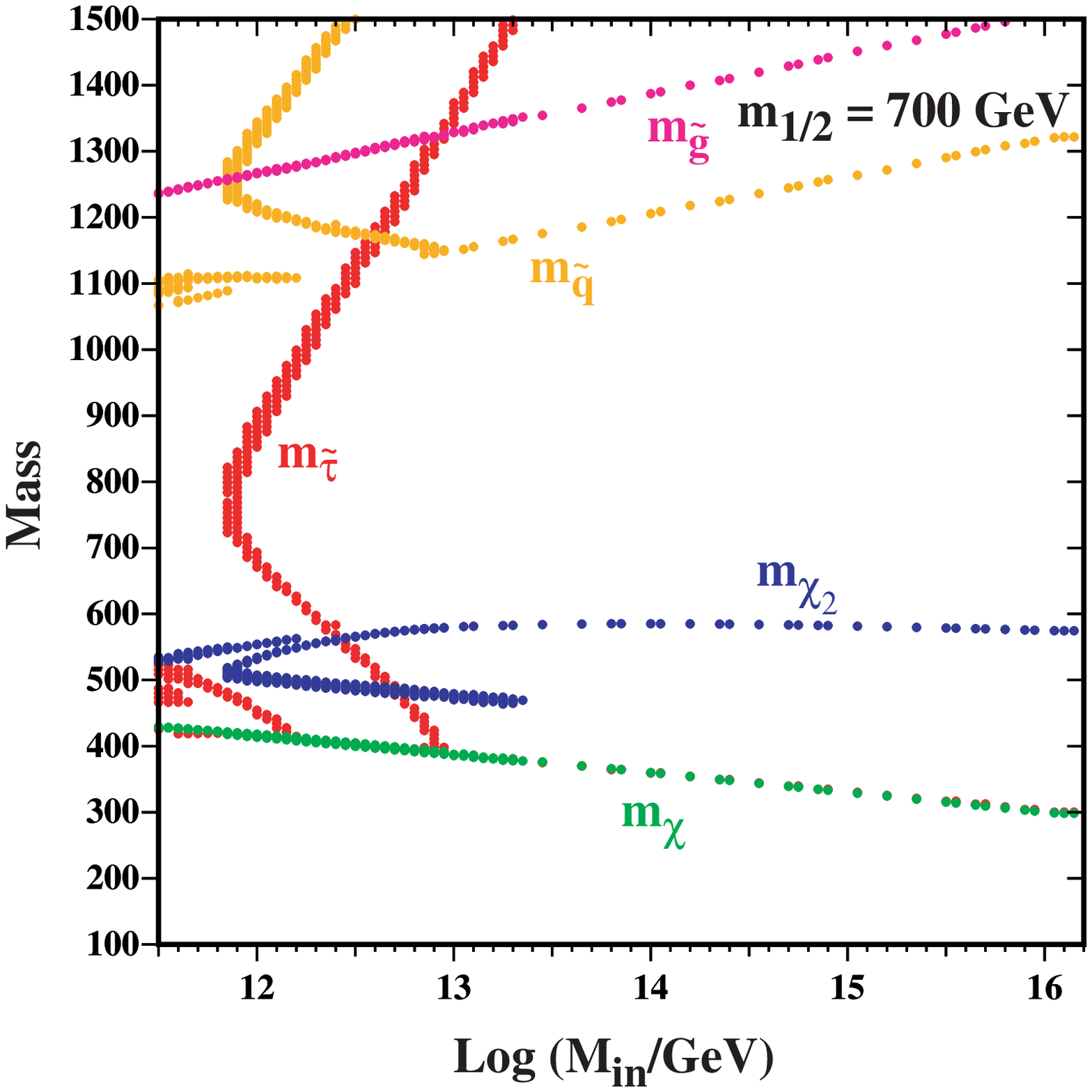,height=8cm}}
\mbox{\epsfig{file=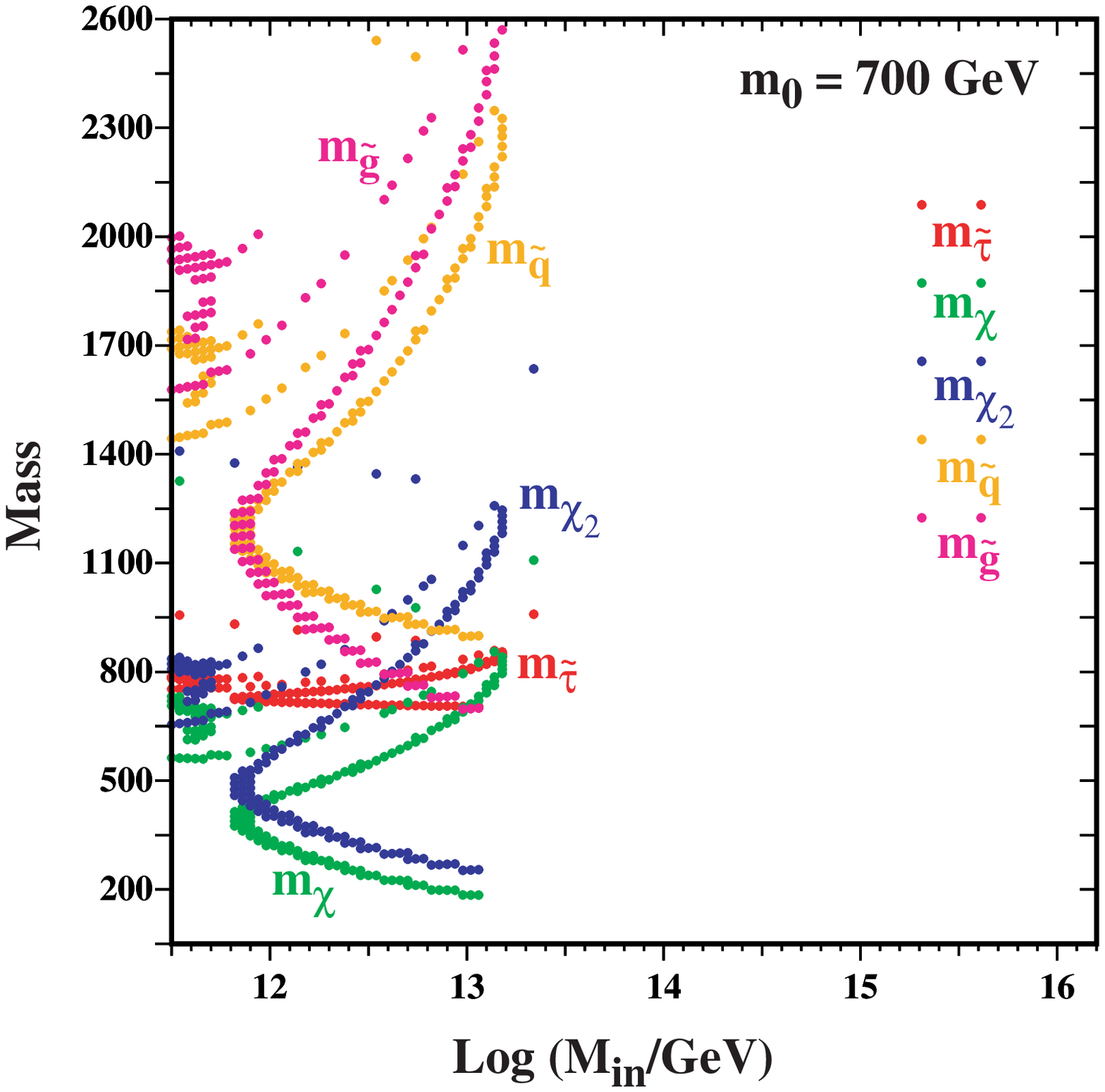,height=8cm}}
\end{center}
\caption{\it
Sparticle masses for sub-GUT CMSSM models chosen to be compatible with the WMAP
relic-density constraint for $\tan \beta = 10, A = 0, \mu > 0$ and (a) $m_{1/2} = 700$~GeV,
(b) $m_0 = 700$~GeV.  For each value of $M_{in}$, we choose (a) $m_0$ and (b)
$m_{1/2}$ so as to respect WMAP, and then plot the corresponding sparticle masses as
functions of $M_{in}$.}
\label{fig:linear2}
\end{figure}

It is clear that the prospects for searches for supersymmetry at the LHC and
other colliders depend on the value of $M_{in}$ assumed. One may also
ask to what extent collider measurements could be used to extract the
value of $M_{in}$, at least within a specific CMSSM framework. These are
complicated issues whose full investigation would extend far beyond the scope of 
this exploratory study. Here we restrict our attention to two specific scans across the
$(m_{1/2}, m_0)$ plane for $\tan \beta = 10$ and $\mu > 0$ as functions of $M_{in}$, 
shown in Fig.~\ref{fig:linear2}. 
In scan (a), we first fix $m_{1/2} = 700$~GeV and then, for each value of $M_{in}$, find
the values(s) of $m_0$ that yield a relic density within the range favoured by WMAP.
Then, for each of these WMAP-compatible choices of $m_0$, we calculate the
masses of some interesting sparticles, namely $\chi, {\tilde \tau_1}, \chi_2, {\tilde q_R}$
and ${\tilde g}$ and finally we plot their dependences on $M_{in}$. 
In scan (b), we instead first fix $m_0 = 700$~GeV, then find,
for each value of $M_{in}$, the value(s) of $m_{1/2}$ yielding the WMAP relic density,
and finally plot the same set of masses as functions of $M_{in}$.

In the case of the first scan at fixed $m_{1/2} = 700$~GeV shown in
Fig.~\ref{fig:linear2}(a), as $M_{in}$ decreases from $2 \times
10^{16}$~GeV towards $10^{13}$~GeV, we see that $m_{\tilde g}$ and
$m_{\tilde q_R}$ decrease gradually, whereas $m_\chi, m_{\tilde \tau_1}$
and $m_{\chi_2}$ increase gradually. The behaviours of $m_{\tilde g}$ and
$m_\chi$ are simply due to their reduced mass renormalizations as $M_{in}$
decreases. In the case of $m_{\tilde \tau_1}$, at large $M_{in}$, one must
choose $m_0$ to lie within the WMAP coannihilation strip, so that the
relic density remains within the allowed range. This requires $m_{\tilde
\tau_1}$ to be only very slightly larger than $m_\chi$~\footnote{For this
reason, the (red) ${\tilde \tau_1}$ points are scarcely visible along the
(green) $\chi$ line.}, so it also increases as $M_{in}$ decreases. In the
case of $m_{\tilde q_R}$, there are effects due to both the reduced mass
renormalization and the WMAP-induced change in $m_0$, the former being
dominant. A new phenomenon appears as $M_{in} \to 10^{13}$~GeV, namely, as
seen in Fig.~\ref{fig:mint}(d), the WMAP strip at small $m_0$ moves away
from the coannihilation limit, and $m_{\tilde \tau_1}$ increases much more
rapidly than $m_\chi$. Also, a new branch of the WMAP strip
appears~\footnote{This branch, associated with the focus point, exists at
larger $M_{in}$ as well, but it does not appear in our scan, because it
only extends to $m_0 = 1500$~GeV.} at large $m_0$, in which $m_{\tilde
\tau_1}$ decreases as $M_{in} \to 10^{12}$~GeV: similar behaviour is
apparent for $m_{\tilde q_R}$. For points in the upper $m_0$ branch, the
$\chi_2$ has a lower mass and is predominantly Higgsino in content,
whereas in the lower $m_0$ branch the $\chi_2$ is mostly wino.  When
$M_{in} \sim 10^{12}$~GeV as shown in Fig.~\ref{fig:mint2}, the two
branches of the WMAP strip merge, as do the two possible values of
$m_{\tilde \tau_1}, m_{\tilde q_R}$ and $m_{\chi_2}$. However, appearing
already at $M_{in}$ slightly larger than $10^{12}$~GeV, we see new,
somewhat lower ranges of allowed values of $m_{\tilde \tau_1}$ and
$m_{\tilde q_R}$ (and higher values of $m_{\chi_2}$), which correspond to
the wedge of allowed $m_0$ values inside the `vee' visible in
Fig.~\ref{fig:mint}(d) for $m_{1/2}$ beyond the rapid-annihilation funnel.
It is apparent that the spectra allowed by WMAP are very sensitive to the
assumed value of $M_{in}$. For example, a determination of the ratio
$m_\chi/m_{\tilde g}$ with an accuracy of 4~\% (which may be possible at
the LHC) would by itself fix $M_{in}$ to within an order of magnitude, in
the restricted set of models considered here.

In the case of the second scan at $m_0 = 700$~GeV, we see in
Fig.~\ref{fig:mint} that due to the Higgs mass bound (we use here the
value of 112 GeV calculated using {\tt FeynHiggs}, so as to account for
theoretical uncertainties), a suitable WMAP strip appears only when
$M_{in} \la 10^{13}$~GeV, and this is reflected in the disappearance of
the sparticle mass lines just above $M_{in} = 10^{13}$~GeV in
Fig.~\ref{fig:linear2}(b). As $M_{in}$ decreases, two of the branches for
each sparticle mass merge. However, there are two other branches, one
appearing near $M_{in} \sim 10^{13}$~GeV and the other closer to $M_{in}
\sim 10^{12}$~GeV. These are due to the appearance of the WMAP-allowed
`vee' seen close to the $m_\chi = m_{\tilde \tau_1}$ line in
Fig.~\ref{fig:mint}(d) {\it et seq.}. In this case, we see that the
WMAP-allowed values of the sparticle masses vary rapidly for $M_{in} \in
(10^{12}, 10^{13})$~GeV. This another example how LHC measurements of
sparticle masses would help fix the magnitude of $M_{in}$ in this
restricted set of models.

\section{Discussion}

We have presented a first exploration of the dependence of the $(m_{1/2},
m_0)$ plane for $\tan \beta = 10, A = 0, \mu > 0$ on the scale $M_{in}$ at
which the input soft supersymmetry-breaking CMSSM mass parameters
$m_{1/2}$ and $m_0$ are assumed to be universal. We have displayed and
explained how the phenomenological, experimental and cosmological
constraints vary with $M_{in}$. In particular, we have shown that the
morphology of the region favoured by the WMAP range of the relic density
changes with $M_{in}$. Specifically, the focus point region at large $m_0$
the coannihilation strip and the rapid-annihilation funnel at large
$m_{1/2}$ approach each other and merge as $M_{in}$ decreases to $\sim
10^{12}$~GeV. Consequently, the values of the sparticle masses that would
be compatible with WMAP depend on $M_{in}$, and measurements at the LHC
may be able to offer some hints about the value of $M_{in}$ within such
sub-GUT CMSSM scenarios.

It is desirable to extend this discussion to other values of the CMSSM
parameters $\tan \beta$ and $A$. It would also be interesting to extend
this analysis to less constrained versions of the MSSM, such as models
with non-universal Higgs masses, and also more constrained versions of the
MSSM motivated by minimal supergravity. It would also be valuable to
extend the brief discussion given here of the corresponding spectra and
the prospects for the LHC and ILC to `measure' indirectly the value of
$M_{in}$. We plan to return to these issues in a future paper.

\section*{Acknowledgments}
\noindent 
The work of K.A.O. and P.S. was supported in part
by DOE grant DE--FG02--94ER--40823.

\end{document}